%% file: Proceeding.tex
\documentclass[a4paper,11pt]{article}
\usepackage{pos}
\usepackage{amsmath}
\usepackage{nicefrac}
\usepackage[utf8]{inputenc}
\usepackage{subcaption}
\usepackage{wrapfig}

\everymath{\displaystyle}
\DeclareMathAlphabet{\mathcal}{OMS}{cmsy}{m}{n}

\title{Two- and three-particle scattering in the (1+1)-dimensional O(3) non-linear sigma model}
\ShortTitle{Two- and three-particle scattering in the O(3) model}

\author*[a]{Jorge Baeza-Ballesteros}
\author[b]{Maxwell T. Hansen}

\affiliation[a]{IFIC, CSIC-Universitat de València, 46980 Paterna, Spain}

\affiliation[b]{School of Physics, University of Edinburgh, Edinburgh EH9 3JZ, UK}

\emailAdd{jorge.baeza@uv.es}
\emailAdd{maxwell.hansen@ed.ac.uk}

\abstract{We study two- and three-particle scattering in the O(3) non-linear sigma model in 1+1 dimensions, focusing on the isospin-1 and isospin-2 channels for two particles, and the isospin-3 channel for three. We perform numerical simulations for four values of the physical volume, each at three lattice spacings, using a three-cluster generalization of the cluster update algorithm, and directly extrapolate the determined finite-volume energies to the continuum at fixed physical volume. Lattice results for two particles are then compared against exact predictions, obtained by combining analytic results for the scattering phase shifts and the (1+1)-dimensional two-particle formalism that relates these to finite-volume energies. Analogous comparisons in the three-particle sector are underway, making use of the three-particle relativistic-field-theory finite-volume formalism. }

\FullConference{%
  The 39th International Symposium on Lattice Field Theory (Lattice2022),\\
  8-13 August, 2022 \\
  Bonn, Germany \\
  Report number: IFIC/22-37
}

\begin{document}
\maketitle

\section{Scattering in finite volume}
\label{sec:intro}

During the last decade, tremendous progress has been made in the determination of multiparticle scattering properties using lattice techniques. This is however a highly non-trivial, since numerical simulations are restricted to finite volumes, and so infinite-volume scattering quantities are not directly accessible. Instead, one can only compute the finite-volume energy spectrum of stationary multiparticle states, that then needs to be related to infinite-volume observables. This is done using so-called quantization conditions.

For two identical spinless particles, a quantization condition was first proposed in a seminal work by M. Lüscher~\cite{Luscher:1986pf, Luscher:1990ux}, and has since been generalized to include any possible two-particle process. It takes the form,
\begin{equation}\label{eq:QC2}
\det[\mathcal{K}_{2}^{-1}+F(P,L)]=0,
\end{equation}
where $\mathcal{K}_2$ is the two-particle infinite-volume $K$-matrix, related to the two-particle scattering phase shift, and $F$ is a finite-volume geometric factor related to power-law finite-volume effects arising from two-particle intermediate states that can go on shell. It depends on the box size, $L$, and the total momentum of the center-of-mass (CM) of the system, $P$.

For three particles, different frameworks have been developed during the last decade: the ``relativistic field theory'' (RFT) approach~\cite{Hansen:2014eka, Hansen:2015zga}, the ``non-relativistic effective field theory'' approach~\cite{Hammer:2017uqm, Hammer:2017kms}, and the ``finite-volume unitarity'' approach~\cite{Mai:2017vot, Mai:2017bge}. All of them have been proved to be equivalent in certain limits~\cite{Blanton:2020jnm}. In this work, we focus on the first of them.

The RFT formalism relates finite-volume quantities to scattering properties in two steps:
\begin{enumerate}
\item Two- and three-particle $K$ matrices are constrained from finite-volume energies using the quantization condition, 
\begin{equation}\label{eq:QC3}
\det[\mathcal{K}_\text{df,3}^{-1}+F_3(\mathcal{K}_2;P,L)]=0.
\end{equation}
Here the three-particle $K$ matrix, $\mathcal{K}_\text{df,3}$, is an intermediate scheme-dependent infinite-volume quantity, while $F_3$ is a geometric factor that depends on $L$ and $P$, but also on the two-particle $K$ matrix. In case of identical spinless particles with a $\mathbb{Z}_2$ symmetry and no two-particle resonances, it takes the form
\begin{equation}\label{eq:F3}
F_3=\frac{1}{3}\tilde{F}+\tilde{F}\frac{1}{\tilde{\mathcal{K}}_2^{-1}-(\tilde{F}+G)}\tilde{F},
\end{equation}
where $\tilde{F}$ and $\tilde{\mathcal{K}}_2$ are modified versions of $F$ and $\mathcal{K}_2$, respectively, and $G$ is a new geometric factor that appears in the derivation of the three-particle quantization condition.
\item Infinite-volume scattering amplitudes are determined from $\mathcal{K}_2$ and $\mathcal{K}_\text{df,3}$ by solving infinite-volume integral equations.
\end{enumerate}

This formalism has been generalized to theories including two-to-three processes~\cite{Briceno:2017tce}, two-particle resonances~\cite{Briceno:2018aml}, multiple subchannels~\cite{Hansen:2020zhy}, and non-identical particles~\cite{Blanton:2020gmf}. In addition, it has been successfully applied to some systems containing pions and kaons at maximal isospin~\cite{Blanton:2019vdk, Hansen:2020otl, Blanton:2021eyf}. However, it has never been used in a non-perturbative integrable theory for which analytic predictions are known in advance (see however Ref.~\cite{Garofalo:2022pux}). In this work, we aim at testing the RFT approach using a solvable model, which will allow us to better understand the intricacies of the formalism.

\section{O(3) non-linear sigma model in 1+1 dimensions}

We have chosen the (1+1)-dimensional O(3) non-linear sigma model as our test bed for the RFT formalism. This model has been widely used as a toy model of QCD due to important qualitative similarities of the two theories. For instance, it was recently used to test a new procedure to compute spectral functions~\cite{Bulava:2021fre}. The model has the following action,
\begin{equation}
S[\sigma]=\frac{\beta}{2}\int\text{d}^2 x\,\partial_\mu\sigma(x)\cdot\partial^\mu\sigma(x),
\end{equation}
where $\beta$ is a dimensionaless coupling constant, and $\sigma$ is a three-component real field of unit length, $\sigma(x)\cdot\sigma(x)=1$. This theory is asymptotically free, and possesses a dynamically generated mass gap, $m$, and a global O(3) symmetry. It thus presents a low-energy particle spectrum consisting of an isospin-1 multiplet.

Furthermore, an exact analytic solution for the two-particle $S$-matrix was proposed in the 1970s based on the properties of unitarity, crossing symmetry and factorization~\cite{Zamolodchikov:1977nu, Zamolodchikov:1978xm}. The latter is a feature of some (1+1)-dimensional theories by which any multiparticle $S$-matrix can be written as products of two-particle $S$-matrices.

\begin{wrapfigure}{r}{0.472\textwidth}
   \centering
   \begin{minipage}{0.458\textwidth}
   \includegraphics[width=1\textwidth,clip]{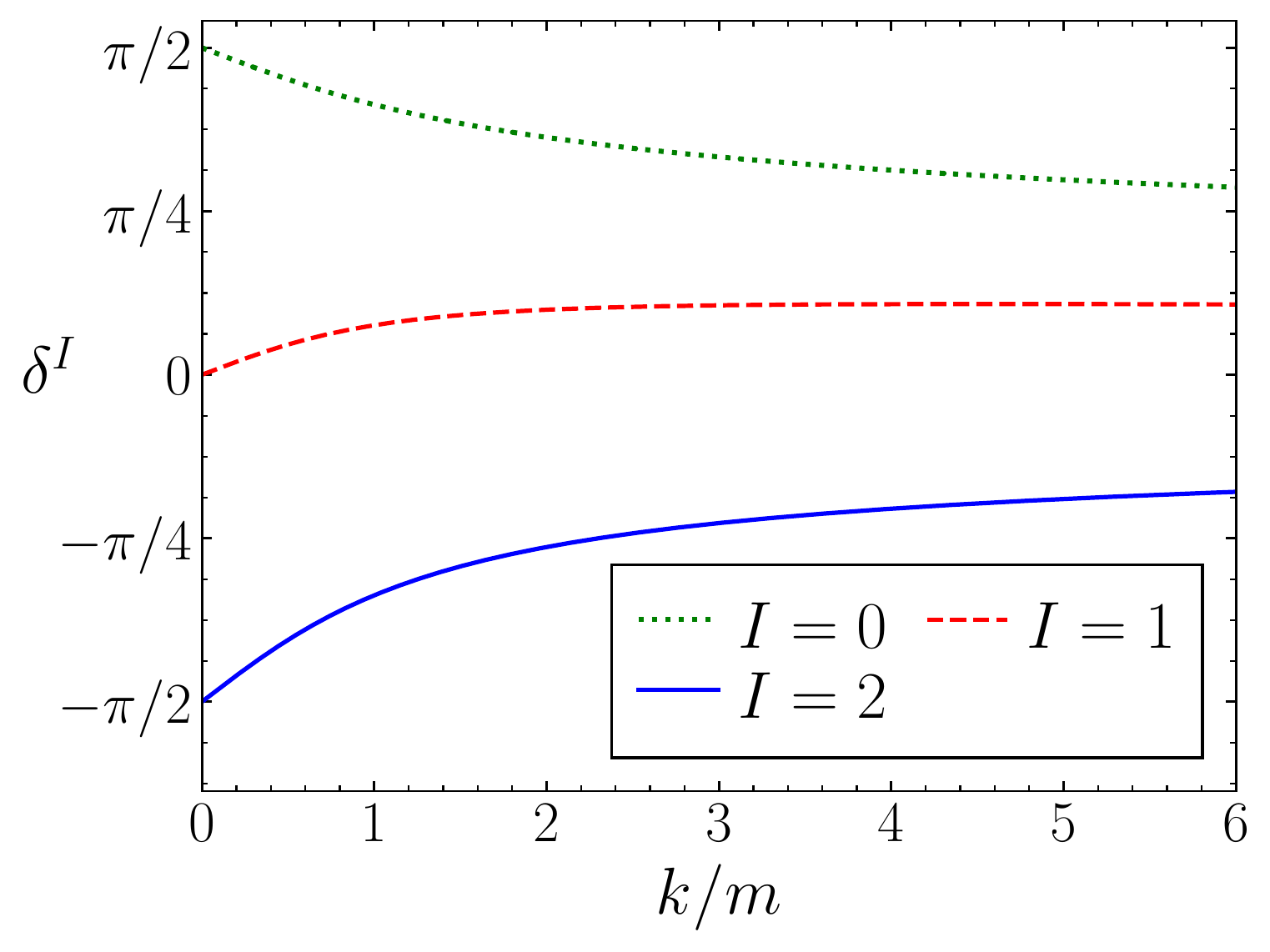}
   \caption{Analytic results for the two-particle scattering phase shifts in all isospin channels.}
   \label{fig:2particlephaseshifts}
   \end{minipage}
\end{wrapfigure}

\begin{figure}[h!]\vspace{-0.15cm}
   \centering
\begin{subfigure}[b]{0.4\linewidth}
\centering%
             \begin{minipage}{0.28\textwidth}
\includegraphics[width=\textwidth]{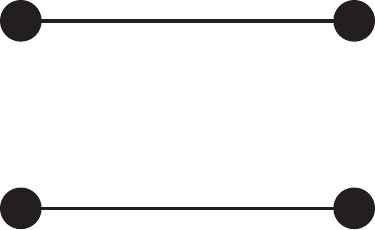} 

       \centering \scalebox{1.2}{$A_2$}
\end{minipage} \hspace{0.37 cm}
\begin{minipage}{0.28\textwidth}
\includegraphics[width=\textwidth]{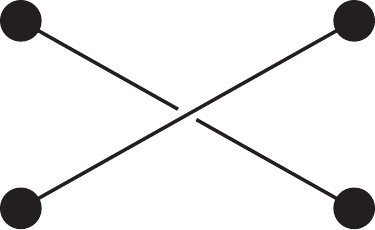} 

        \centering\scalebox{1.2}{$ A_3$}
\end{minipage}\vspace{0.45cm}

\caption{Two-particle contractions}
\label{fig:contraction2part}
\end{subfigure}
\begin{subfigure}[b]{0.55\linewidth}
\centering%
             $\begin{array}{c}
\includegraphics[width=0.2\textwidth]{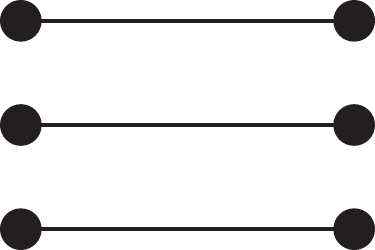}  \hspace{0.5 cm}
\includegraphics[width=0.2\textwidth]{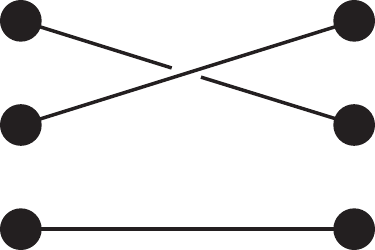}  \hspace{0.5 cm}
\includegraphics[width=0.2\textwidth]{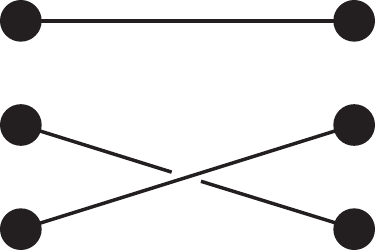}  \vspace{0.3cm}\\

\includegraphics[width=0.2\textwidth]{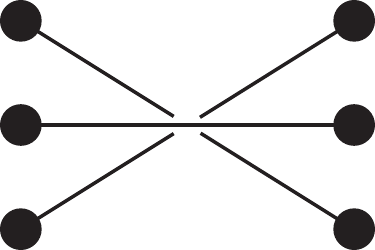}  \hspace{0.5 cm}
\includegraphics[width=0.2\textwidth]{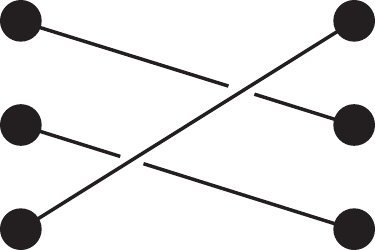}  \hspace{0.5 cm}
\includegraphics[width=0.2\textwidth]{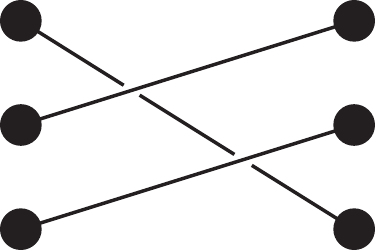}  \vspace{0.cm}\\
\end{array}$
\caption{Three-particle contractions}
\label{fig:contraction3part}
\end{subfigure}
 \caption{Diagramatic representation of those Wick contractions relevant for the two-particle isospin-1 and isospin-2 channels, and the three-particle isospin-2 and isospin-3 channels.}\vspace{-0.2cm}
   \label{fig:contractions}
\end{figure}

In this talk, we report our progress on the study of two- and three-particle scattering for those channels that contain no vacuum contractions. Notably, the scattering channels in the O(3) model coincide with those of two-flavor QCD, which is unique among other factorizable theories. For two particles, there are three different scattering channels corresponding to different isospin values, $I$. Using the exact result for the two-particle $S$-matrix, it is possible to determine the corresponding scattering phase shifts, which are shown in Fig.~\ref{fig:2particlephaseshifts}. In this work, we focus on the $I=1$ and $I=2$ channels, for which the lattice correlation functions can be computed as linear combinations of two Wick contractions (see Fig~\ref{fig:contraction2part}),
\begin{equation}
C_{I=1}=A_2-A_3,\quad\quad\quad C_{I=2}=A_2+A_3.
\end{equation}

In the three-particle sector, the combination of three isospin-1 fields leads to seven irreducible representations of the O(3) group, which organize in four different scattering channels. The multiplicity of each channel corresponds to the number of possible two-particle subchannels that can appear. This seven-fold multiplicity also exists for three-pion states in QCD, as recently discussed in Ref.~\cite{Hansen:2020zhy}. The full scope of this work will be to consider the isospin-2 and isospin-3 channels, for which the lattice correlation functions are linear combinations of six non-vacuum contractions (see Fig.~\ref{fig:contraction3part}). Note that, for the former channel, one in general requires a $2\times 2$ matrix of correlators for each momenta combination due to the presence of two two-particle subchannels. In this talk we restrict our attention to the isospin-3 channel.

Two- and three-particle correlation functions can be used to determine the finite-volume energy spectrum. This then needs to be related to infinite-volume quantities using (1+1)-dimensional versions of the quantization conditions. These are simpler than their (3+1)-dimensional counterparts, due to the absence of angular momentum in one spatial dimension. 

For two particles, the quantization condition in Eq.~(\ref{eq:QC2}) simplifies to the following algebraic equation, which was worked out in Refs.~\cite{Guo:2013vsa, Briceno:2021aiw},
\begin{equation}\label{eq:2particleQC2dim}
\cot\delta(k)=-\frac{1}{2}\left\{\cot\left[\frac{L\gamma(k+\omega_k\beta)}{2}\right]+\cot\left[\frac{L\gamma(k-\omega_k\beta)}{2}\right]\right\},
\end{equation}
where $k$ and $\omega_k=\sqrt{m^2+k^2}$ are the relative momenta and the single-particle energy in the CM frame, respectively, $\gamma$ and $\beta$ are boost factors to that frame, and $\delta(k)$ is the two-particle phase shift. 

\begin{figure}[b!]\vspace{-0.18cm}
   \centering
   \begin{subfigure}[t]{0.45\linewidth}
\centering%
\includegraphics[width=1\textwidth,clip]{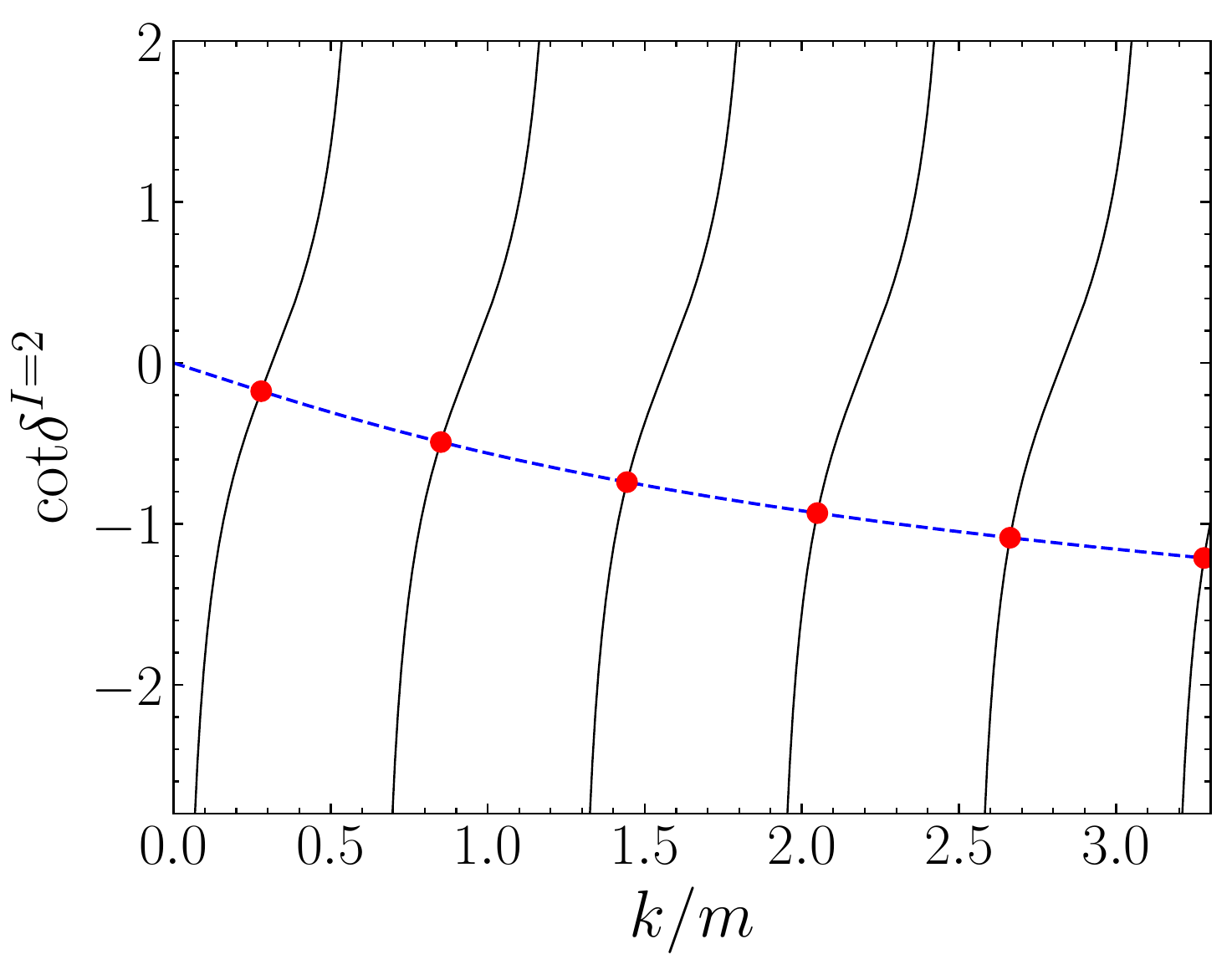}
\caption{Finite-volume energy levels for fixed $mL=10$ (red dots). The dashed blue (solid black) line represents the left (right) hand side of Eq.~(\ref{eq:2particleQC2dim}), for the $I=2$ channel.  }
\label{fig:mL10energies}
\end{subfigure}\hspace{0.4cm}
\begin{subfigure}[t]{0.45\linewidth}
\centering%
\includegraphics[width=1\textwidth,clip]{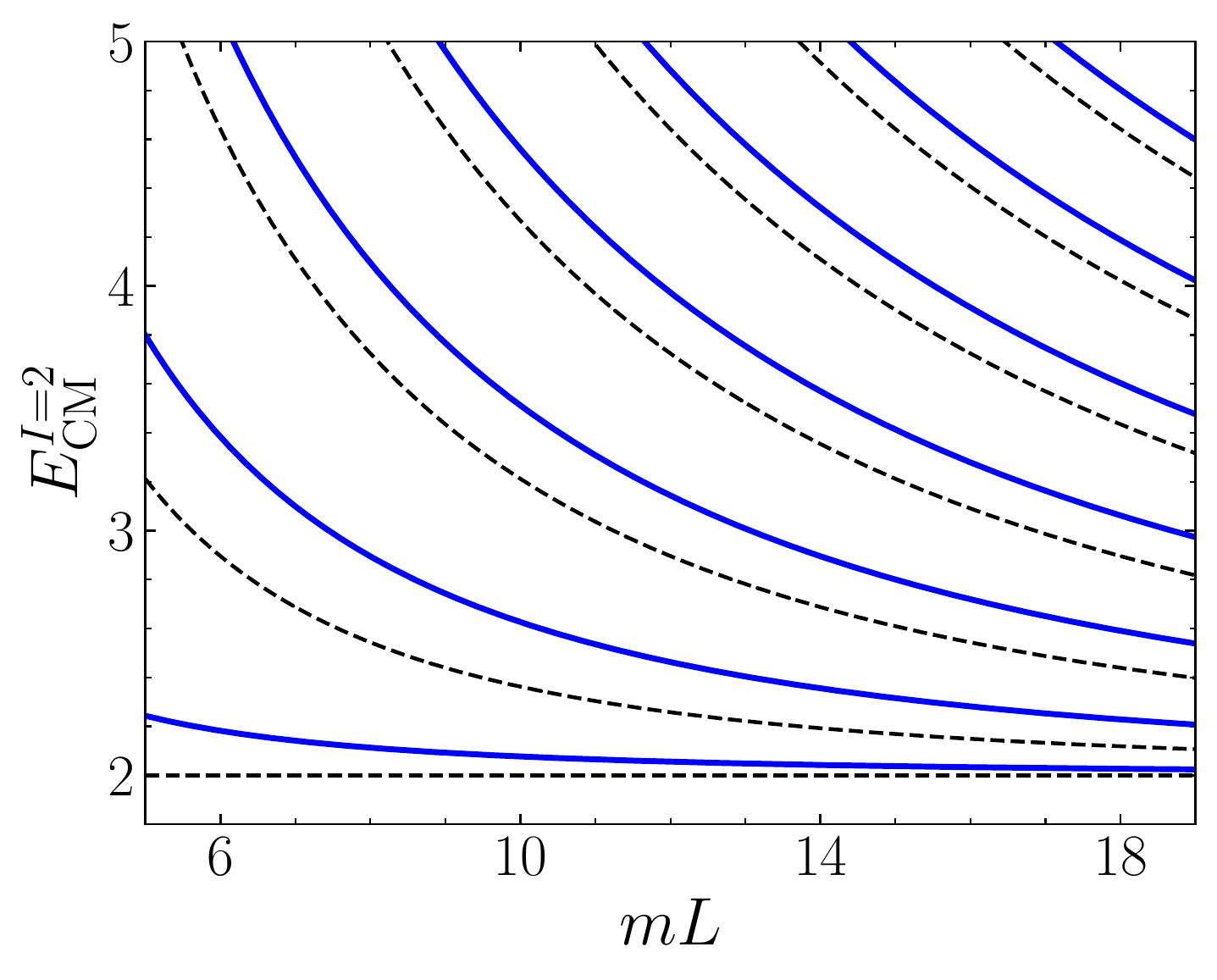}
\caption{Two-particle finite-volume energies for the $I=2$ channel (blue solid) together with the non-interacting energy levels (black dashed).  }
\label{fig:VolumeDepentEnergies}
\end{subfigure}
   \caption{Analytical predictions for the finite-volume energies for the two-particle isospin-2 channel in the rest frame ($P=0$). For a fixed value of $mL$, they are determined solving Eq.~(\ref{eq:2particleQC2dim}) (left panel). Varying $mL$ allows to obtain the full volume-dependent energy spectrum (right panel).}
   \label{fig:example2particleenergies}
\end{figure}

Eq.~(\ref{eq:2particleQC2dim}) can be used together with analytical results of the two-particle scattering phase shifts to obtain exact predictions of the finite-volume energies. This is depicted in Fig.~\ref{fig:example2particleenergies} for the $I=2$ channel in the $P=0$ frame. First, one fixes $L$ and $P$ and solves Eq.~(\ref{eq:2particleQC2dim}) for the energy levels (left panel). This process is then repeated for varying $L$ to obtain the full volume-dependent finite-volume energies (right panel).

For the RFT three-particle quantization condition, Eq.~(\ref{eq:QC3}), the full extent of simplifications in 1+1 dimensions is still under investigation. The quantity $\tilde{F}$ becomes a straightforward function, closely related to the right-hand side of Eq.~(\ref{eq:2particleQC2dim}), and $G$ also simplifies due to the lack of angular momentum. We are currently working on a determination of the finite-volume energies assuming $\mathcal{K}_\text{df,3}=0$, and are also trying to analytically predict $\mathcal K_{\text{df},3}$, in order to give a full RFT-based prediction of the finite-volume energies to compare to our numerical results.

\section{Lattice simulations and the cluster algorithm}

Turning to the numerical side of the project, we have determined finite-volume energy levels for two and three particles using numerical simulations of a lattice-discretized model. We have employed the standard discretized lattice action
\begin{equation}\label{eq:O3actionlattice}
S[\sigma]=-\beta\sum_{x}\sum_\mu\sigma(x)\cdot\sigma(x+a\hat{\mu}),
\end{equation}
where $a$ is the lattice spacing, $x=(\tau, \boldsymbol{x})$ denotes the lattice site, $\hat{\mu}$ is a unit vector in the direction $\mu$, and the two sums run over all the lattice, and over space and time directions, respectively. We have considered periodic boundary condition in both space and time.

For the generation of configurations and the evaluation of $n$-point functions we have used the cluster algorithm. It is a collective algorithm applicable to spin systems that overcomes critical slowing down  and vastly improves the signal-to-noise ratio. It was first proposed in Ref.~\cite{Wolff:1988uh} for a single cluster, and then generalized to a two-cluster update in Ref.~\cite{Luscher:1990ck}. In this work, we have further generalized it to three clusters so as to be able to study three-particle scattering processes.

The single-cluster algorithm works as follows: First, a random unit vector $r\in S^2$ is uniformly drawn from the unit sphere, and a ``seed'' lattice site is randomly chosen. From this site, a cluster $C$ is grown: for each site $x$ belonging to $C$, each of the non-cluster neighbours $y$ are considered and added to the cluster with probability
\begin{equation}
p_\text{add}=1-\exp[\min\{-2\beta\sigma_r(x)\sigma_r(y), 0\}],
\end{equation} 
where $\sigma_r(x)=\sigma(x)\cdot r$. This is repeated for all the newly added sites until all neighbours have been considered. Then a new configuration is obtained by updating all spins in $C$,
\begin{equation}
\sigma(x)\rightarrow\sigma(x)-2\sigma_r(x)r.
\end{equation}

The cluster can then be used to obtain cluster-improved estimators of the quantites of interest. For example, for the two-point function the spins in the cluster are first projected to definite momentum $p$,
\begin{equation}
\sigma_r(\tau,p)=\sum_{\boldsymbol{x}\in C}\sigma(x)\text{e}^{ip\boldsymbol{x}},
\end{equation}
and are then used to compute the desired result,
\begin{equation}
C_\text{2pt}(\tau,p)\propto\langle\sigma_r(\tau, p)\sigma_r^*(0,p)\rangle,
\end{equation}
where $\langle\cdot\rangle$ indicates that time translation invariance is used to average over all equivalent time separations. This quantity then needs to be averaged over several consecutive updates, in order to obtain a sensible result.

Because neither the single- nor the double-cluster algorithms allow us to compute three-particle correlation functions, we have had to generalize to three clusters. In this case, three random orthonormal vectors are drawn from the unit sphere, $r, u, v\in S^2$, and three ``seed'' lattice sites are randomly chosen. Each of them is used to grow an independent cluster, $C_r$, $C_u$ and $C_v$, respectively, using the same procedure as in the single-cluster case. Note that, owing to the orthogonality of the vectors, the growth of the clusters do not interfere with each other. Finally, the three clusters are updated and used to measure cluster-improved estimators of the contractions represented in Fig.~\ref{fig:contractions}. For example, a four-point correlation function can be computed using two of the clusters, and a six-point one using all of them,
\begin{align}
\def\svgwidth{0.11\textwidth}
        \vcenter{\hbox{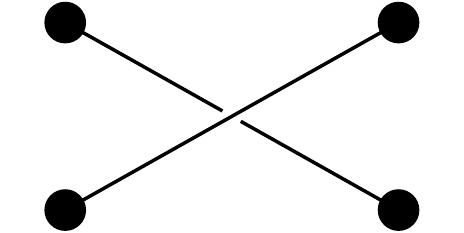 }} \hspace{.3cm}&\propto\langle\sigma_r(\tau, q_2)\sigma_u(\tau,q_1)\sigma_u^*(0,p_2)\sigma_r^*(0,p_1)\rangle,\\[0.35cm]
\def\svgwidth{0.11\textwidth}
        \vcenter{\hbox{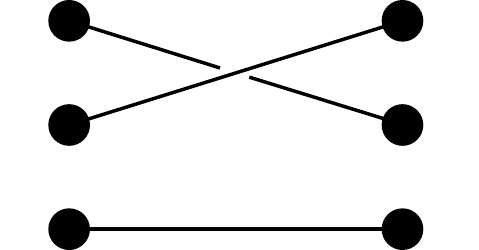 }} \hspace{.3cm}&\propto\langle\sigma_r(\tau,q_3)\sigma_u(\tau,q_2)\sigma_v(\tau,q_1)\sigma_r^*(0,p_3)\sigma_v^*(0,p_2)\sigma_u^*(0,p_1)\rangle,
\end{align}
 where $p_i$ and $q_i$ denote the momenta of particles in the initial and final states, respectively, and $\sigma_u$, $\sigma_v$ are defined analogously to $\sigma_r$.

\begin{figure}[h!]
   \centering
\includegraphics[width=0.5\textwidth,clip]{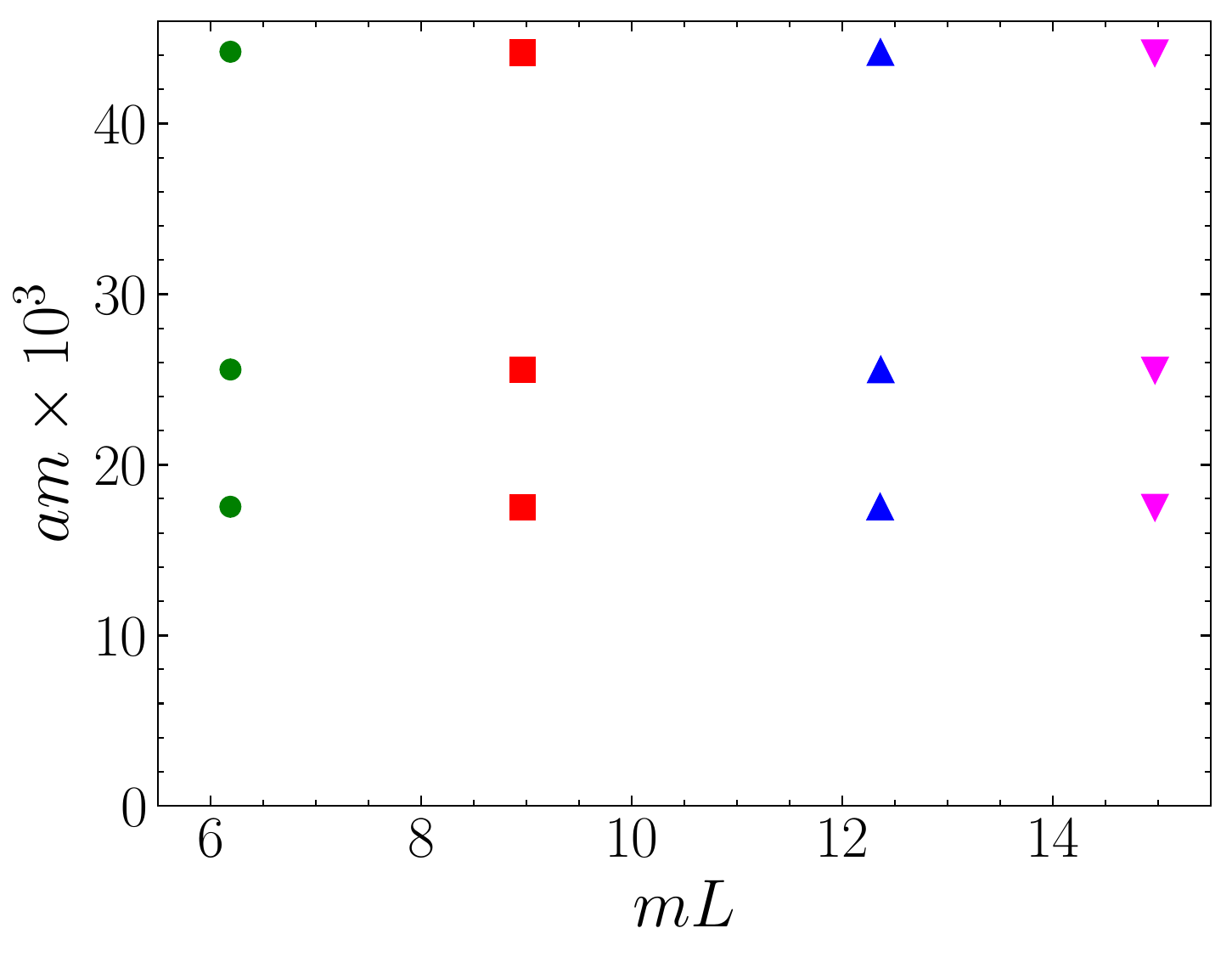}
   \caption{Summary of the ensembles used for the simulations. All have $mT\sim20.4$ to avoid thermal effects.}
   \label{fig:summaryensembles}
\end{figure}

Numerical simulations have been performed using the \texttt{o3\_cluster} code, an early version of which was provided to us by J. Bulava~\cite{Bulava:2022}. We have generated 12 ensembles, with four values of $mL$, each at three lattice spacings. The physical volume, $mL$, and time extent, $mT$, of the lattice have been finely tuned so as to be able to directly extrapolate the finite-volume energy levels to the continuum. In particular, we have used $mL> 6$ so that exponentially suppressed finite-volume corrections can be neglected, and set $mT\sim 20$ to be safe of thermal effects. A summary of the ensembles is presented in Fig.~\ref{fig:summaryensembles}.

For each ensemble, we have generated $N_\text{rep}=1024$, 512 and 256 replicas, for the coarser, intermediate and finest ensembles, respectively. Each of them has been thermalized for $12.8\times10^6$ three-cluster updates, and the correlation functions of interest have then been averaged over 4$\times10^6$, 2$\times10^6$ and 1$\times10^6$ consecutive updates, in this same order. In all cases, we have studied four different values of total spatial momentum in the finite-volume frame. To reliably extract the finite-volume energies, we have used a basis of operators that exhausts all individual-particle momentum combinations for the targeted range of energies.

\section{Results for the two- and three particle energy spectra}

\begin{figure}[h!]
   \centering
   \begin{subfigure}[t]{0.45\linewidth}
\centering%
\includegraphics[width=1\textwidth,clip]{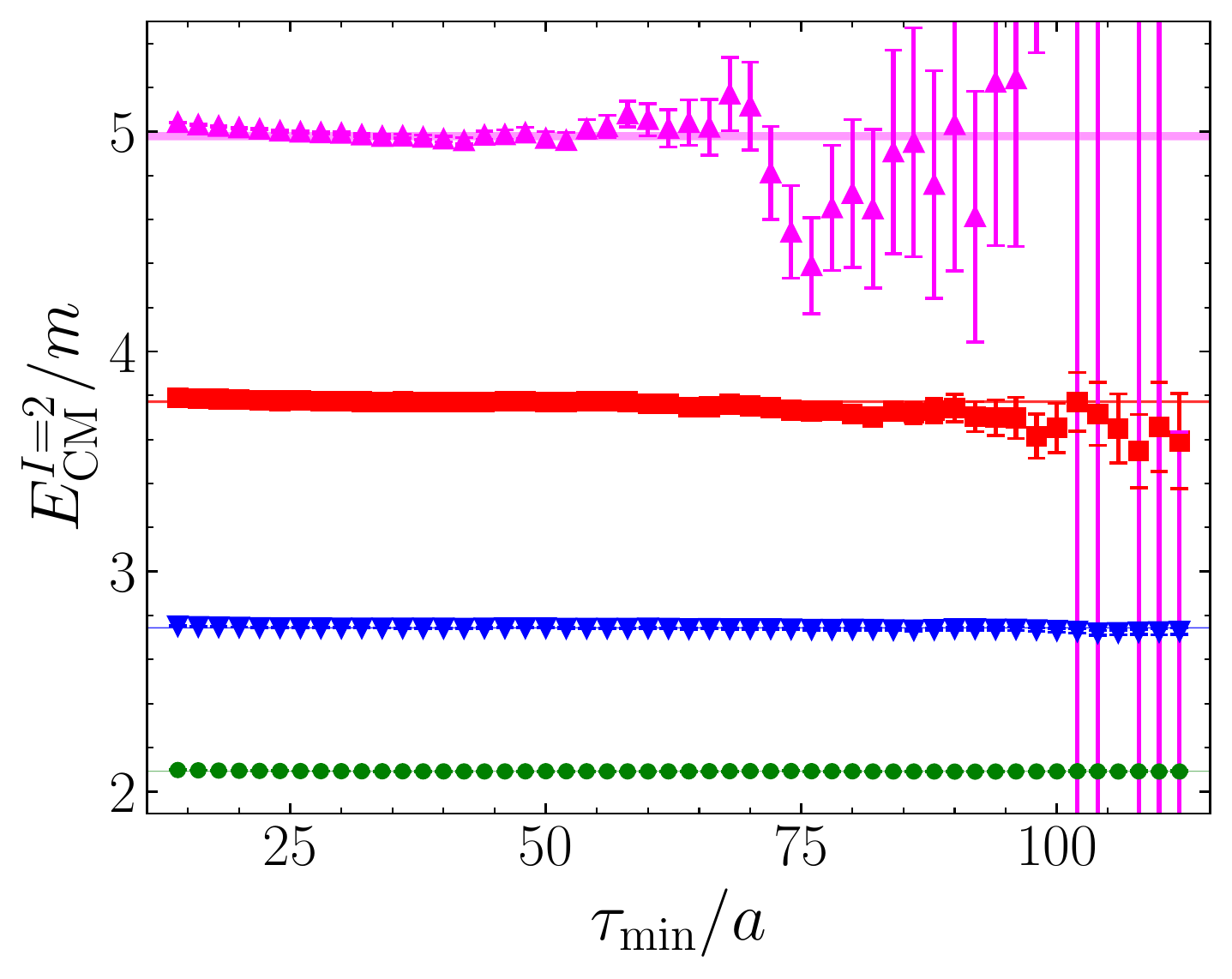}
\caption{Two-particle $I=2$ channel.  }
\end{subfigure}\hspace{0.4cm}
\begin{subfigure}[t]{0.45\linewidth}
\centering%
\includegraphics[width=1\textwidth,clip]{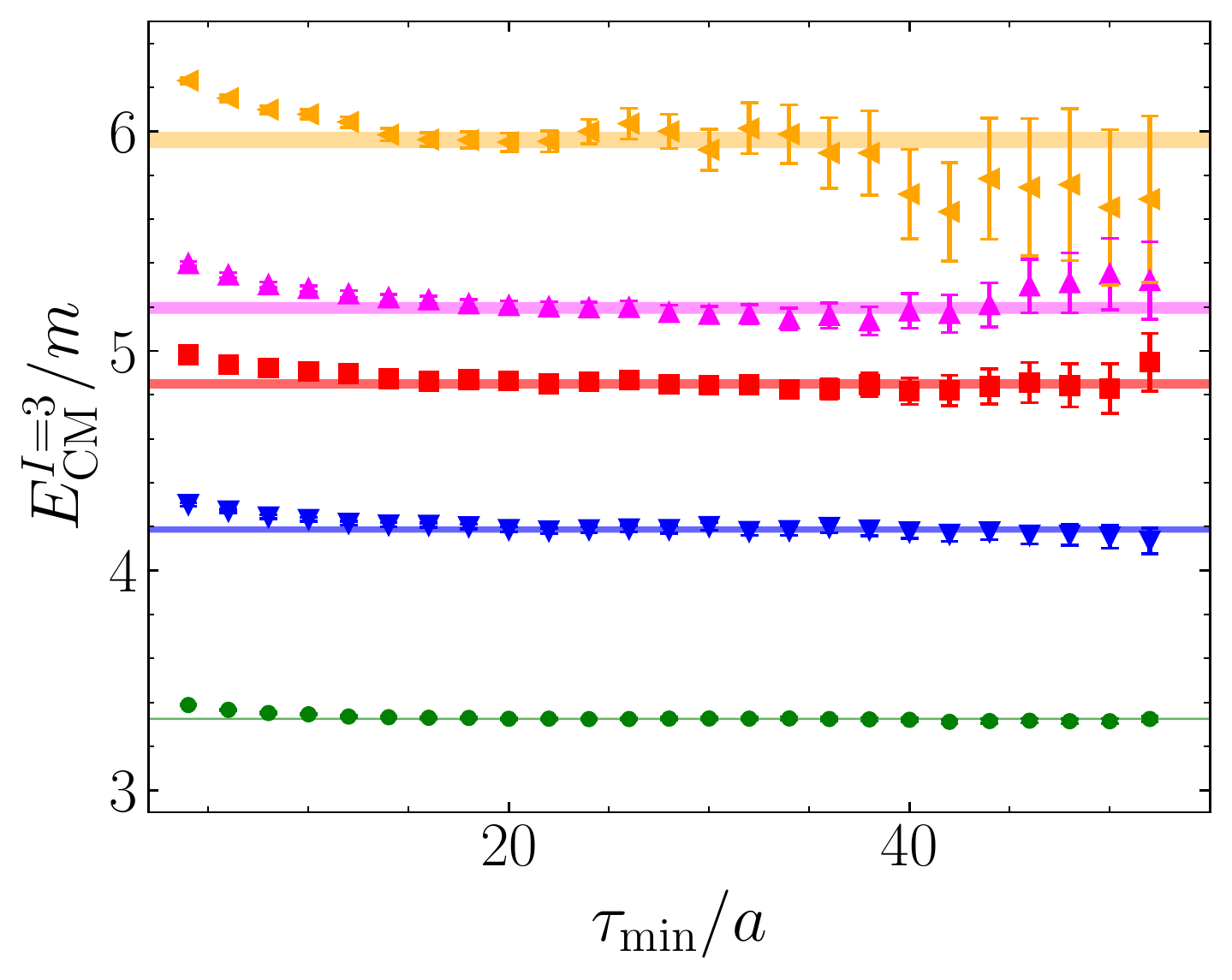}
\caption{Three-particle $I=3$ channel. }
\end{subfigure}
   \caption{Rest-frame energy spectra obtained for different fit ranges, $[\tau_\text{min},\tau_\text{max}]$, with varying $\tau_\text{min}$ and fixed $\tau_\text{max}$, for an ensemble with $mL\sim 9$ and $am \sim 0.025$. Final results (horizontal bands) are extracted from the plateaux.}
   \label{fig:plateau}
\end{figure}

We have computed two- and three-particle correlation functions for the channels of interest, and used them to solve a generalized eigenvalue problem. The eigenvalues have been fitted to a single exponential for different fit ranges, and the finite-volume energies, $E_n$, have been determined by ensuring the results showed a plateau with respect to the fit range chosen. Two examples are presented in Fig.~\ref{fig:plateau} for two- and a three-particle cases.

\begin{figure}[h!]
   \centering
   \begin{subfigure}[t]{0.45\linewidth}
\centering%
\includegraphics[width=1\textwidth,clip]{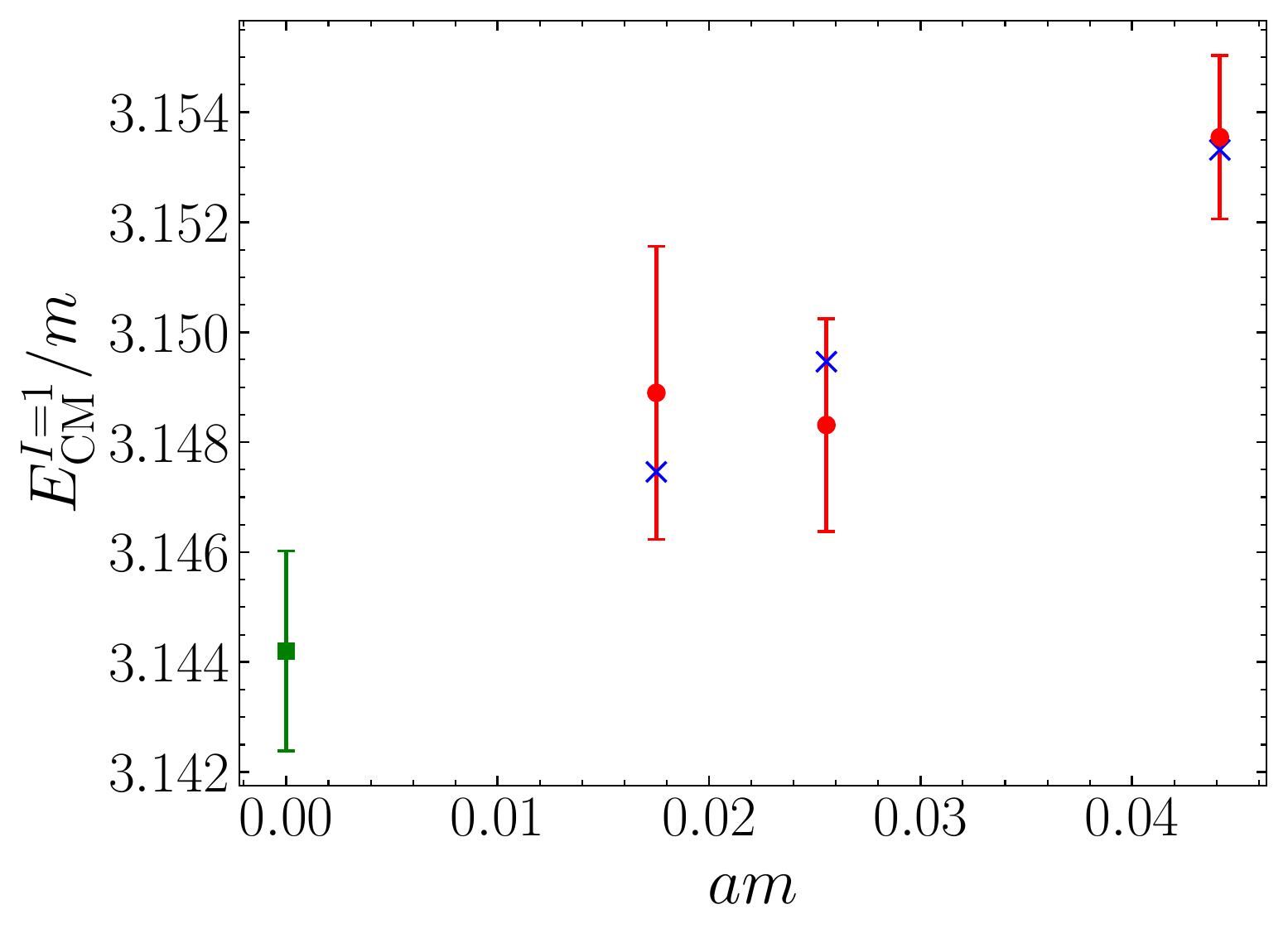}
\caption{$I=1$ channel for a $mL\sim12$ ensemble. }
\end{subfigure}
   \begin{subfigure}[t]{0.45\linewidth}
\centering%
\includegraphics[width=1\textwidth,clip]{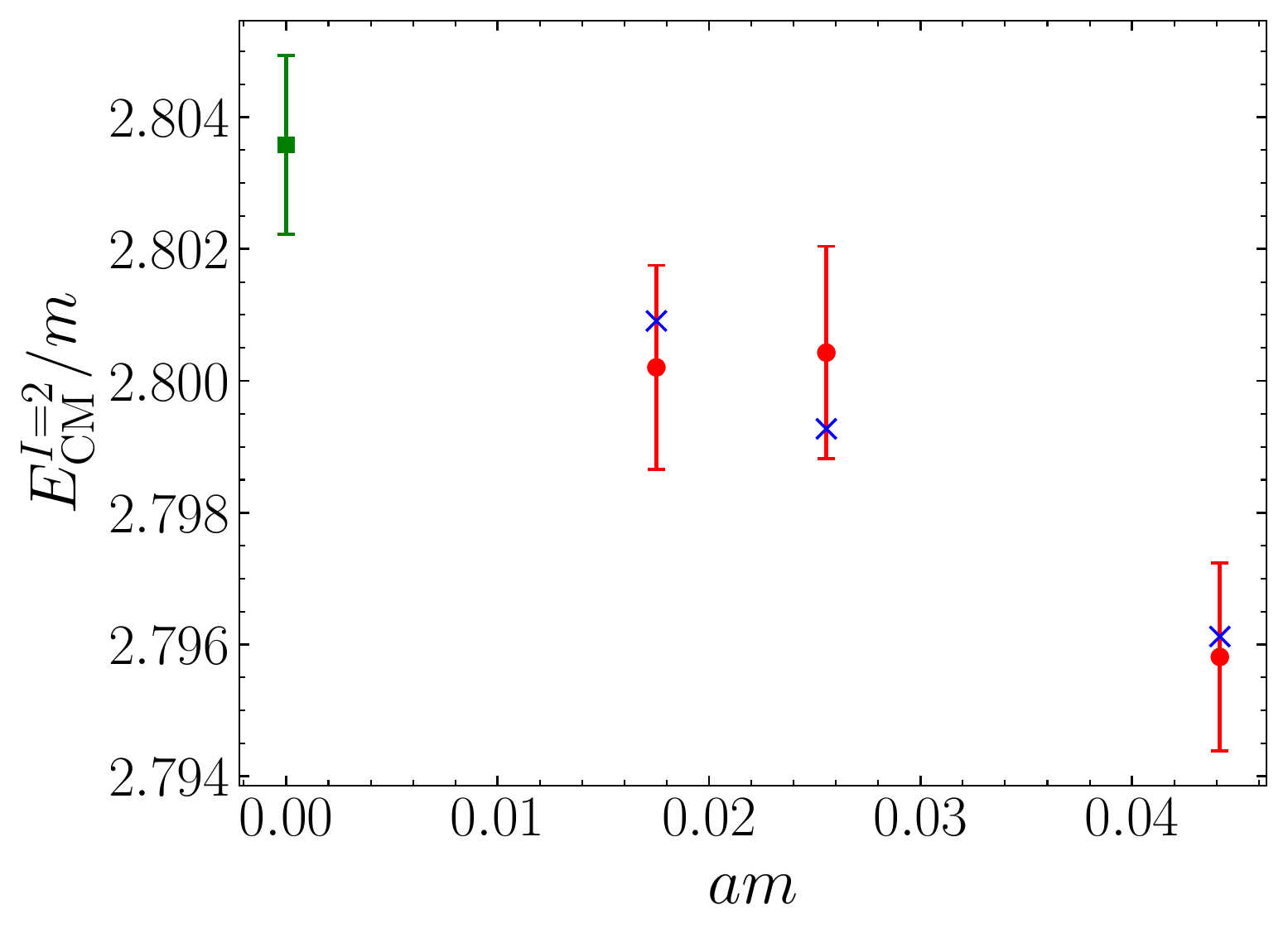}
\caption{$I=2$ channel for a $mL\sim15$ ensemble.  }
\end{subfigure}\hspace{0.4cm}

   \caption{Continuum extrapolation of the finite-volume energy levels. Blue crosses represent the best-fit predictions to Eq.~(\ref{eq:continuumlimit}) up to $k=2$ terms.}
   \label{fig:continuumextrapolation}
\end{figure}

The energy levels have directly been extrapolated to the continuum. It is well known that the O(3) model presents large logarithmic discretization effects. These have been analytically studied in detail~\cite{Balog:2009yj, Balog:2009np}, and on-shell quantities, such as the finite-volume energies, are known to present the following asymptotic behavior,
\begin{equation}\label{eq:continuumlimit}
Q(am)=Q(0)+C(am)^2\beta^3\left[1+\sum_{k=1}^\infty c_k\beta^{-k}\right] + \mathcal{O}(a^4),
\end{equation}
where $C$ and $c_{k\geq 3}$ depend on the particular observable, while $c_1$ and $c_2$ are fixed for any on-shell quantity and can be computed from two- and three-loop integrals, respectively. For the standard discretized action, Eq.~(\ref{eq:O3actionlattice}), they are
\begin{equation}
c_1=-1.1386...,\quad\quad\quad c_2=-0.4881.
\end{equation}
In this work we have considered $k=1,2$ terms, and fitted $Q(0)$ and $C$ for each energy level. Two examples of extrapolations are shown in Fig.~\ref{fig:continuumextrapolation}.

\begin{figure}[b!]
   \centering
  \begin{subfigure}[t]{0.8\linewidth}
\centering%
\includegraphics[width=1\textwidth,clip]{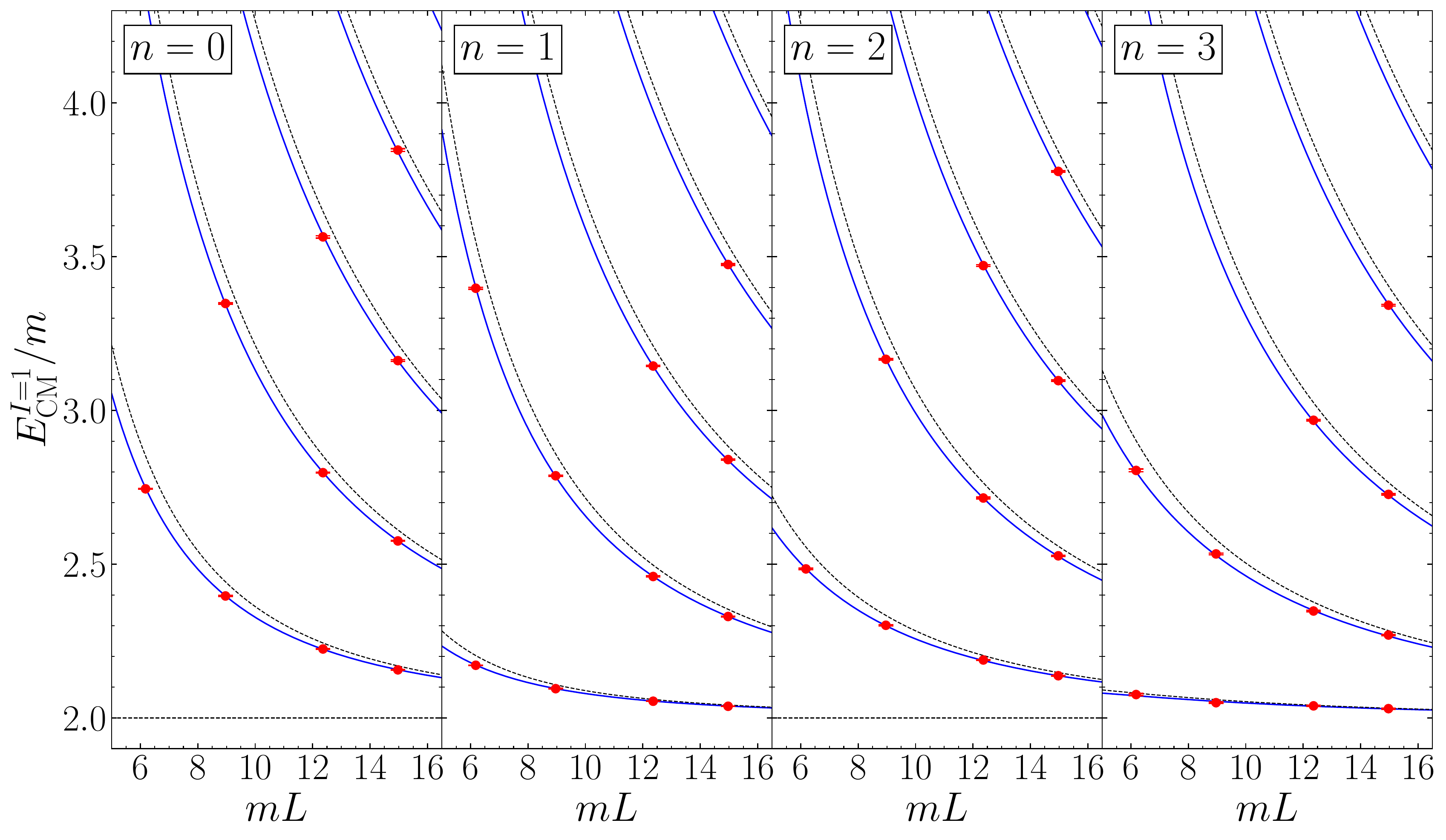}
\caption{$I=1$ channel. }
\end{subfigure}\vspace{0.5cm}
   \begin{subfigure}[t]{0.8\linewidth}
\centering%
\includegraphics[width=1\textwidth,clip]{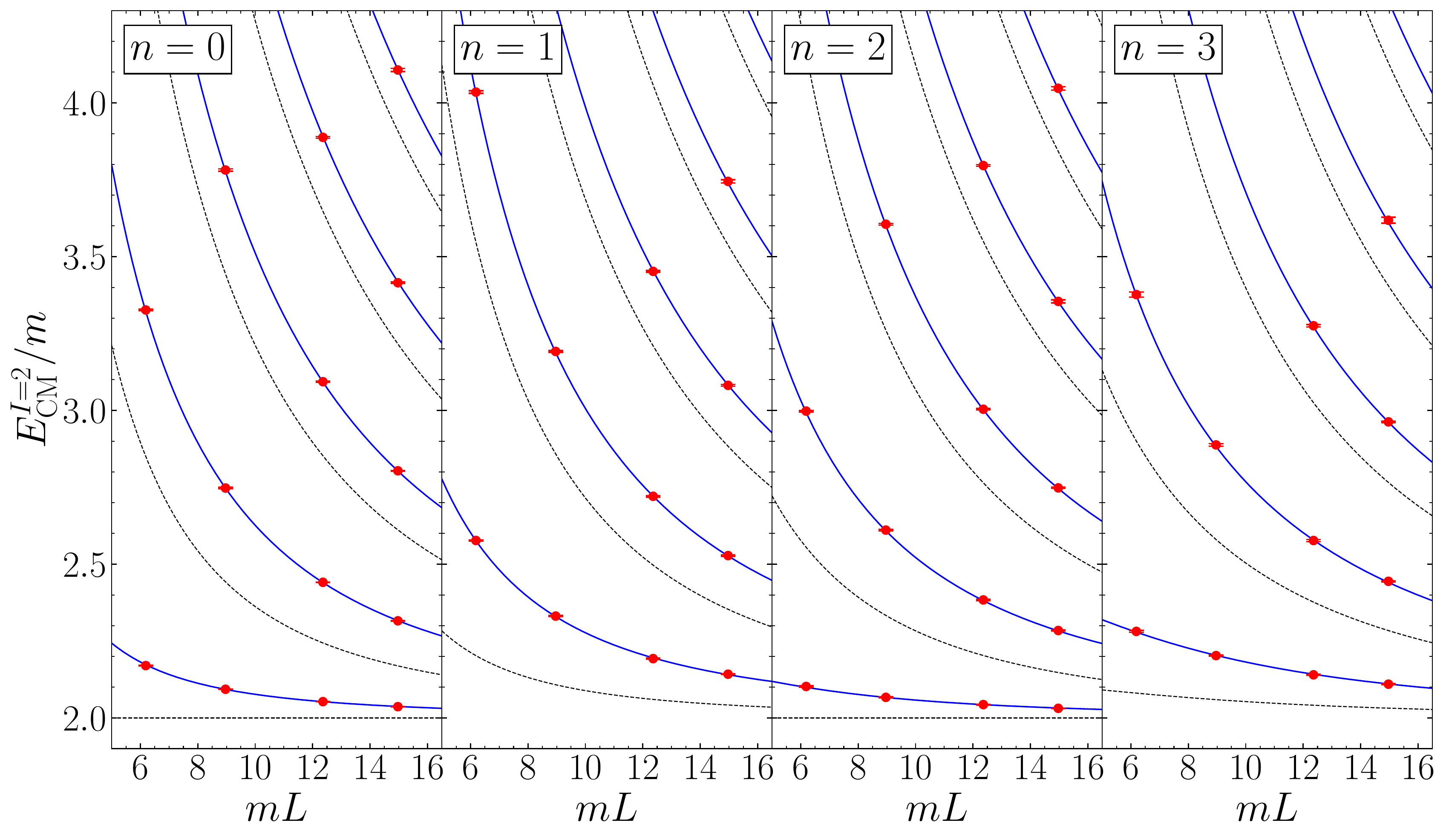}
\caption{$I=2$ channel.  }\vspace{0.4cm}
\end{subfigure}

   \caption{Preliminary results for continuum-extrapolated two-particle finite-volume energies. Lattice results (red points) are represented together with analytic predictions (blue solid lines) and free energies (black dashed lines). Each panel corresponds to a different momentum frame, with $P=\frac{2\pi}{L}n$. }
   \label{fig:twoparticleenergies}
\end{figure}

Preliminary results for the two-particle finite-volume energy levels are shown in Fig.~\ref{fig:twoparticleenergies} for the $I=1$ and $I=2$ channels. We compare them to the exact predictions computed using Eq.~(\ref{eq:2particleQC2dim}) (blue solid lines). Two-particle free energies (black dashed lines) are also represented for completeness. We observe reasonable agreement between the numerical and the analytical results, which we quantify by computing the $\chi^2$ between both,
\begin{equation}
\begin{array}{rcl}
\chi^2/\text{dof}=1.26, & \quad\quad\quad\quad\quad\quad (I=1 \text{ channel}), \\
\chi^2/\text{dof}=1.34, & \quad\quad\quad\quad\quad\quad (I=2 \text{ channel}).
\end{array}
\end{equation}
We stress these are not the results of a fit to lattice data, but a comparison between them and analytical expectations. Also note that no inelastic threshold is considered, as no particle production can occur in a factorizable theory.

\begin{figure}[h!]
   \centering
\includegraphics[width=0.8\textwidth,clip]{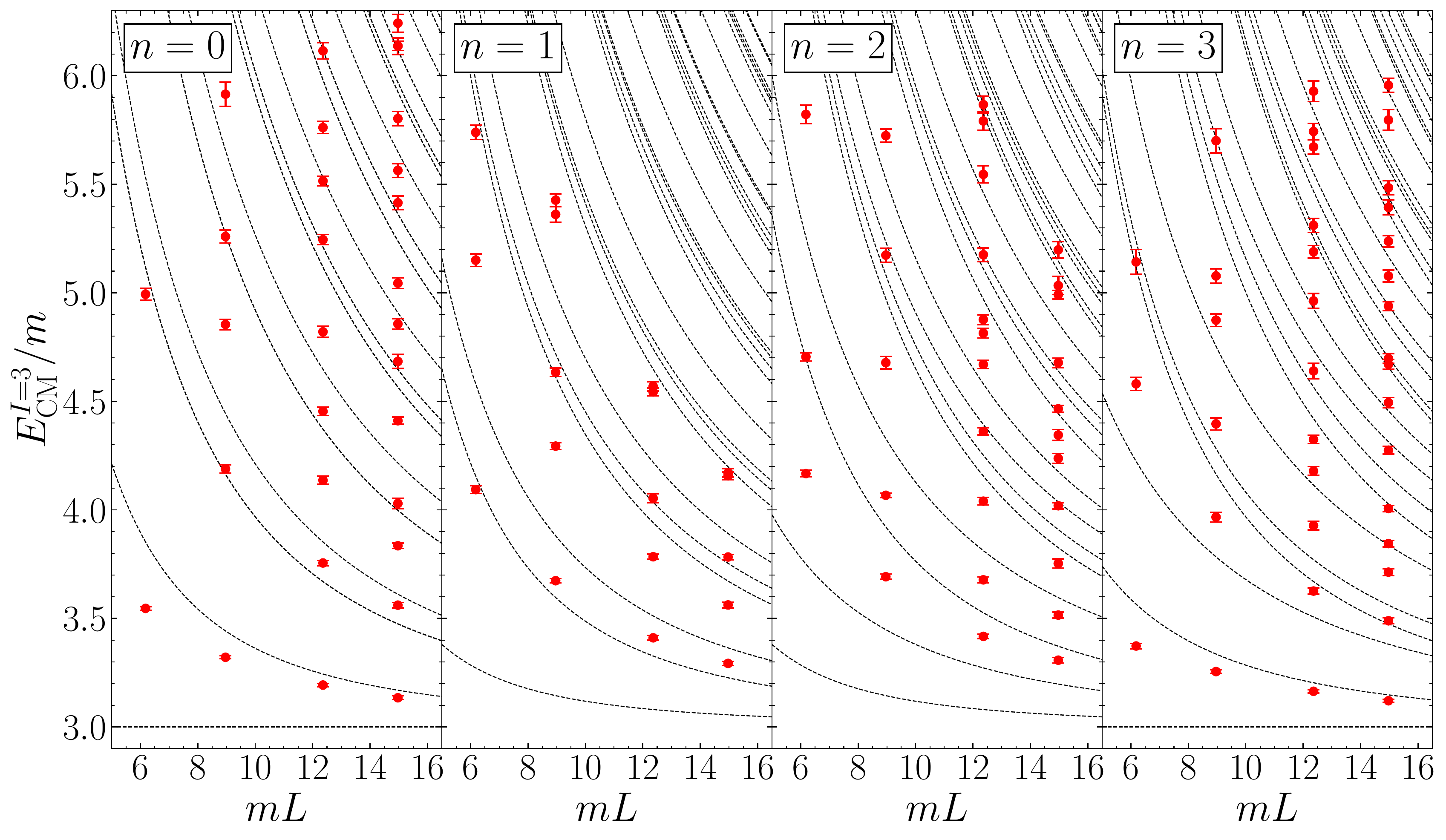}
   \caption{Preliminary results for the three-particle finite-volume energies in the $I=3$ channel at $am\sim0.017$, together with the free energies (black dashed lines). Each panel corresponds to a different momentum frame, with $P=\frac{2\pi}{L}n$.}
   \label{fig:threeparticleenergiesI3}\vspace{-0.1cm}
\end{figure}

Finally, a computation of three-particle energies in the $I=2$ and $I=3$ channels is underway. Preliminary results for the latter one corresponding to the finest ensembles are presented in Fig.~\ref{fig:threeparticleenergiesI3}, together with the free energy levels. At the moment, we are working on a prediction of the interacting energies under the assumption $\mathcal{K}_\text{df,3}=0$, and also trying to obtain an analytic determination of $\mathcal{K}_\text{df,3}$.

\section{Summary and outlook}

In this talk, we have reported the current status of our study of two- and three-particle scattering in the integrable (1+1)-dimensional O(3) non-linear sigma model, with which we aim at testing the three-particle RFT finite-volume formalism. We have implemented a three-cluster algorithm and used it to compute two- and three-particle finite-volume energies, finding very good agreement, at the sub-percent level, between lattice and analytical expectations in the two-particle case. 

In addition to cross-checking and finalizing all aspects of the calculation, we are currently working on extending the extraction of finite-volume energies to include the three-particle $I=2$ channel. The analytic determination of the three-particle energies based on the RFT formalism is also ongoing. We are considering the predictions for $\mathcal{K}_\text{df,3}=0$ (of interest for benchmarking the sensitivity that the finite-volume energies might have to this quantity), as well as the more complicated case of predicting $\mathcal{K}_\text{df,3}=0$ directly from the integrable theory. The isospin-2 three-particle channel is especially interesting as it makes use of the formalism with coupled two-particle subsystems derived in Ref.~\cite{Hansen:2020zhy}.


\section{Acknowledgments}

We thank John Bulava for useful discussion and for providing us with some code to simulate the O(3) model. JBB also thanks Pilar Hernández for helpful suggestions regarding the talk. 

This work was performed in part under the Project HPC-EUROPA3 (INFRAIA-2016-1-730897), with the support of the EC Research Innovation Action under the H2020 Programme. JBB gratefully acknowledges the hospitality of the Higgs Centre for Theoretical Physics at the University of Edinburgh for its hospitality during his stay during which this project was started. JBB is supported by the Spanish grant FPU19/04326 of the spanish MU. Additionally, JBB receives support from the Generalitat Valenciana grant PROMETEO/2019/083, the Spanish AEI project PID2020-113644GB-I00/AEI/10.13039/501100011033, and the European project H2020-MSCA-ITN-2019//860881-HIDDeN. MTH is supported in part by the UK STFC grant ST/P000630/1, and by the UKRI Future Leader Fellowship MR/T019956/1. This work used resources from the ARCHER2 UK National Supercomputing Service. JBB wants to thank EPCC at the University of Edinburgh for the technical support provided.\vspace{-0.1cm}

\setlength{\bibsep}{1.3\itemsep}
\bibliographystyle{apsrev4-1.bst}
\bibliography{bibtexref.bib}

\end{document}

%% file: Figures/A3color.pdf_tex
\begingroup%
  \makeatletter%
  \providecommand\color[2][]{%
    \errmessage{(Inkscape) Color is used for the text in Inkscape, but the package 'color.sty' is not loaded}%
    \renewcommand\color[2][]{}%
  }%
  \providecommand\transparent[1]{%
    \errmessage{(Inkscape) Transparency is used (non-zero) for the text in Inkscape, but the package 'transparent.sty' is not loaded}%
    \renewcommand\transparent[1]{}%
  }%
  \providecommand\rotatebox[2]{#2}%
  \newcommand*\fsize{\dimexpr\f@size pt\relax}%
  \newcommand*\lineheight[1]{\fontsize{\fsize}{#1\fsize}\selectfont}%
  \ifx\svgwidth\undefined%
    \setlength{\unitlength}{133.56659459bp}%
    \ifx\svgscale\undefined%
      \relax%
    \else%
      \setlength{\unitlength}{\unitlength * \real{\svgscale}}%
    \fi%
  \else%
    \setlength{\unitlength}{\svgwidth}%
  \fi%
  \global\let\svgwidth\undefined%
  \global\let\svgscale\undefined%
  \makeatother%
  \begin{picture}(1,0.50162242)%
    \lineheight{1}%
    \setlength\tabcolsep{0pt}%
    \put(0,0){\includegraphics[width=\unitlength,page=1]{Figures/A3color.pdf}}%
    \put(0.95376164,0.41473827){\color[rgb]{0.,0,0}\makebox(0,0)[lt]{\lineheight{1.25}\smash{\begin{tabular}[t]{l}$q_1$\end{tabular}}}}%
    \put(0.95376164,0.01044557){\color[rgb]{0.,0.,0}\makebox(0,0)[lt]{\lineheight{1.25}\smash{\begin{tabular}[t]{l}$q_2$\end{tabular}}}}%
    \put(-0.21,0.41473827){\color[rgb]{0.,0.,0}\makebox(0,0)[lt]{\lineheight{1.25}\smash{\begin{tabular}[t]{l}$p_1$\end{tabular}}}}%
    \put(-0.21,0.01044557){\color[rgb]{0.,0,0}\makebox(0,0)[lt]{\lineheight{1.25}\smash{\begin{tabular}[t]{l}$p_2$\end{tabular}}}}%
  \end{picture}%
\endgroup%

%% file: Figures/B2color.pdf_tex
\begingroup%
  \makeatletter%
  \providecommand\color[2][]{%
    \errmessage{(Inkscape) Color is used for the text in Inkscape, but the package 'color.sty' is not loaded}%
    \renewcommand\color[2][]{}%
  }%
  \providecommand\transparent[1]{%
    \errmessage{(Inkscape) Transparency is used (non-zero) for the text in Inkscape, but the package 'transparent.sty' is not loaded}%
    \renewcommand\transparent[1]{}%
  }%
  \providecommand\rotatebox[2]{#2}%
  \newcommand*\fsize{\dimexpr\f@size pt\relax}%
  \newcommand*\lineheight[1]{\fontsize{\fsize}{#1\fsize}\selectfont}%
  \ifx\svgwidth\undefined%
    \setlength{\unitlength}{137.71588688bp}%
    \ifx\svgscale\undefined%
      \relax%
    \else%
      \setlength{\unitlength}{\unitlength * \real{\svgscale}}%
    \fi%
  \else%
    \setlength{\unitlength}{\svgwidth}%
  \fi%
  \global\let\svgwidth\undefined%
  \global\let\svgscale\undefined%
  \makeatother%
  \begin{picture}(1,0.52281549)%
    \lineheight{1}%
    \setlength\tabcolsep{0pt}%
    \put(0,0){\includegraphics[width=\unitlength,page=1]{Figures/B2color.pdf}}%
    \put(-0.2,0.24849145){\color[rgb]{0.,0,0}\makebox(0,0)[lt]{\lineheight{1.25}\smash{\begin{tabular}[t]{l}$p_2$\end{tabular}}}}%
    \put(0.95376164,0.24849145){\color[rgb]{0.,0.,0}\makebox(0,0)[lt]{\lineheight{1.25}\smash{\begin{tabular}[t]{l}$q_2$\end{tabular}}}}%
    \put(-0.2,0.46633124){\color[rgb]{0.,0.,0}\makebox(0,0)[lt]{\lineheight{1.25}\smash{\begin{tabular}[t]{l}$p_1$\end{tabular}}}}%
    \put(0.95376164,0.46633124){\color[rgb]{0.,0,0}\makebox(0,0)[lt]{\lineheight{1.25}\smash{\begin{tabular}[t]{l}$q_1$\end{tabular}}}}%
    \put(-0.2,0.02997092){\color[rgb]{0.,0.,0}\makebox(0,0)[lt]{\lineheight{1.25}\smash{\begin{tabular}[t]{l}$p_3$\end{tabular}}}}%
    \put(0.95240014,0.02997092){\color[rgb]{0.,0.,0}\makebox(0,0)[lt]{\lineheight{1.25}\smash{\begin{tabular}[t]{l}$q_3$\end{tabular}}}}%
  \end{picture}%
\endgroup%

%% file: Proceeding.bbl
\begin{thebibliography}{27}%
\makeatletter
\providecommand \@ifxundefined [1]{%
 \@ifx{#1\undefined}
}%
\providecommand \@ifnum [1]{%
 \ifnum #1\expandafter \@firstoftwo
 \else \expandafter \@secondoftwo
 \fi
}%
\providecommand \@ifx [1]{%
 \ifx #1\expandafter \@firstoftwo
 \else \expandafter \@secondoftwo
 \fi
}%
\providecommand \natexlab [1]{#1}%
\providecommand \enquote  [1]{``#1''}%
\providecommand \bibnamefont  [1]{#1}%
\providecommand \bibfnamefont [1]{#1}%
\providecommand \citenamefont [1]{#1}%
\providecommand \href@noop [0]{\@secondoftwo}%
\providecommand \href [0]{\begingroup \@sanitize@url \@href}%
\providecommand \@href[1]{\@@startlink{#1}\@@href}%
\providecommand \@@href[1]{\endgroup#1\@@endlink}%
\providecommand \@sanitize@url [0]{\catcode `\\12\catcode `\$12\catcode
  `\&12\catcode `\#12\catcode `\^12\catcode `\_12\catcode `\%12\relax}%
\providecommand \@@startlink[1]{}%
\providecommand \@@endlink[0]{}%
\providecommand \url  [0]{\begingroup\@sanitize@url \@url }%
\providecommand \@url [1]{\endgroup\@href {#1}{\urlprefix }}%
\providecommand \urlprefix  [0]{URL }%
\providecommand \Eprint [0]{\href }%
\providecommand \doibase [0]{http://dx.doi.org/}%
\providecommand \selectlanguage [0]{\@gobble}%
\providecommand \bibinfo  [0]{\@secondoftwo}%
\providecommand \bibfield  [0]{\@secondoftwo}%
\providecommand \translation [1]{[#1]}%
\providecommand \BibitemOpen [0]{}%
\providecommand \bibitemStop [0]{}%
\providecommand \bibitemNoStop [0]{.\EOS\space}%
\providecommand \EOS [0]{\spacefactor3000\relax}%
\providecommand \BibitemShut  [1]{\csname bibitem#1\endcsname}%
\let\auto@bib@innerbib\@empty
\bibitem [{\citenamefont {Luscher}(1986)}]{Luscher:1986pf}%
  \BibitemOpen
  \bibfield  {author} {\bibinfo {author} {\bibfnamefont {M.}~\bibnamefont
  {Luscher}},\ }\href {\doibase 10.1007/BF01211097} {\bibfield  {journal}
  {\bibinfo  {journal} {Commun. Math. Phys.}\ }\textbf {\bibinfo {volume}
  {105}},\ \bibinfo {pages} {153} (\bibinfo {year} {1986})}\BibitemShut
  {NoStop}%
\bibitem [{\citenamefont {Luscher}(1991)}]{Luscher:1990ux}%
  \BibitemOpen
  \bibfield  {author} {\bibinfo {author} {\bibfnamefont {M.}~\bibnamefont
  {Luscher}},\ }\href {\doibase 10.1016/0550-3213(91)90366-6} {\bibfield
  {journal} {\bibinfo  {journal} {Nucl. Phys. B}\ }\textbf {\bibinfo {volume}
  {354}},\ \bibinfo {pages} {531} (\bibinfo {year} {1991})}\BibitemShut
  {NoStop}%
\bibitem [{\citenamefont {Hansen}\ and\ \citenamefont
  {Sharpe}(2014)}]{Hansen:2014eka}%
  \BibitemOpen
  \bibfield  {author} {\bibinfo {author} {\bibfnamefont {M.~T.}\ \bibnamefont
  {Hansen}}\ and\ \bibinfo {author} {\bibfnamefont {S.~R.}\ \bibnamefont
  {Sharpe}},\ }\href {\doibase 10.1103/PhysRevD.90.116003} {\bibfield
  {journal} {\bibinfo  {journal} {Phys. Rev. D}\ }\textbf {\bibinfo {volume}
  {90}},\ \bibinfo {pages} {116003} (\bibinfo {year} {2014})},\ \Eprint
  {http://arxiv.org/abs/1408.5933}{arXiv:1408.5933 [hep-lat]}\BibitemShut
  {NoStop}%
\bibitem [{\citenamefont {Hansen}\ and\ \citenamefont
  {Sharpe}(2015)}]{Hansen:2015zga}%
  \BibitemOpen
  \bibfield  {author} {\bibinfo {author} {\bibfnamefont {M.~T.}\ \bibnamefont
  {Hansen}}\ and\ \bibinfo {author} {\bibfnamefont {S.~R.}\ \bibnamefont
  {Sharpe}},\ }\href {\doibase 10.1103/PhysRevD.92.114509} {\bibfield
  {journal} {\bibinfo  {journal} {Phys. Rev. D}\ }\textbf {\bibinfo {volume}
  {92}},\ \bibinfo {pages} {114509} (\bibinfo {year} {2015})},\ \Eprint
  {http://arxiv.org/abs/1504.04248}{arXiv:1504.04248 [hep-lat]}\BibitemShut
  {NoStop}%
\bibitem [{\citenamefont {Hammer}\ \emph
  {et~al.}(2017{\natexlab{a}})\citenamefont {Hammer}, \citenamefont {Pang},\
  and\ \citenamefont {Rusetsky}}]{Hammer:2017uqm}%
  \BibitemOpen
  \bibfield  {author} {\bibinfo {author} {\bibfnamefont {H.-W.}\ \bibnamefont
  {Hammer}}, \bibinfo {author} {\bibfnamefont {J.-Y.}\ \bibnamefont {Pang}}, \
  and\ \bibinfo {author} {\bibfnamefont {A.}~\bibnamefont {Rusetsky}},\ }\href
  {\doibase 10.1007/JHEP09(2017)109} {\bibfield  {journal} {\bibinfo  {journal}
  {JHEP}\ }\textbf {\bibinfo {volume} {09}},\ \bibinfo {pages} {109} (\bibinfo
  {year} {2017}{\natexlab{a}})},\ \Eprint
  {http://arxiv.org/abs/1706.07700}{arXiv:1706.07700 [hep-lat]}\BibitemShut
  {NoStop}%
\bibitem [{\citenamefont {Hammer}\ \emph
  {et~al.}(2017{\natexlab{b}})\citenamefont {Hammer}, \citenamefont {Pang},\
  and\ \citenamefont {Rusetsky}}]{Hammer:2017kms}%
  \BibitemOpen
  \bibfield  {author} {\bibinfo {author} {\bibfnamefont {H.~W.}\ \bibnamefont
  {Hammer}}, \bibinfo {author} {\bibfnamefont {J.~Y.}\ \bibnamefont {Pang}}, \
  and\ \bibinfo {author} {\bibfnamefont {A.}~\bibnamefont {Rusetsky}},\ }\href
  {\doibase 10.1007/JHEP10(2017)115} {\bibfield  {journal} {\bibinfo  {journal}
  {JHEP}\ }\textbf {\bibinfo {volume} {10}},\ \bibinfo {pages} {115} (\bibinfo
  {year} {2017}{\natexlab{b}})},\ \Eprint
  {http://arxiv.org/abs/1707.02176}{arXiv:1707.02176 [hep-lat]}\BibitemShut
  {NoStop}%
\bibitem [{\citenamefont {Mai}\ \emph {et~al.}(2017)\citenamefont {Mai},
  \citenamefont {Hu}, \citenamefont {Doring}, \citenamefont {Pilloni},\ and\
  \citenamefont {Szczepaniak}}]{Mai:2017vot}%
  \BibitemOpen
  \bibfield  {author} {\bibinfo {author} {\bibfnamefont {M.}~\bibnamefont
  {Mai}}, \bibinfo {author} {\bibfnamefont {B.}~\bibnamefont {Hu}}, \bibinfo
  {author} {\bibfnamefont {M.}~\bibnamefont {Doring}}, \bibinfo {author}
  {\bibfnamefont {A.}~\bibnamefont {Pilloni}}, \ and\ \bibinfo {author}
  {\bibfnamefont {A.}~\bibnamefont {Szczepaniak}},\ }\href {\doibase
  10.1140/epja/i2017-12368-4} {\bibfield  {journal} {\bibinfo  {journal} {Eur.
  Phys. J. A}\ }\textbf {\bibinfo {volume} {53}},\ \bibinfo {pages} {177}
  (\bibinfo {year} {2017})},\ \Eprint
  {http://arxiv.org/abs/1706.06118}{arXiv:1706.06118 [nucl-th]}\BibitemShut
  {NoStop}%
\bibitem [{\citenamefont {Mai}\ and\ \citenamefont
  {D\"oring}(2017)}]{Mai:2017bge}%
  \BibitemOpen
  \bibfield  {author} {\bibinfo {author} {\bibfnamefont {M.}~\bibnamefont
  {Mai}}\ and\ \bibinfo {author} {\bibfnamefont {M.}~\bibnamefont {D\"oring}},\
  }\href {\doibase 10.1140/epja/i2017-12440-1} {\bibfield  {journal} {\bibinfo
  {journal} {Eur. Phys. J. A}\ }\textbf {\bibinfo {volume} {53}},\ \bibinfo
  {pages} {240} (\bibinfo {year} {2017})},\ \Eprint
  {http://arxiv.org/abs/1709.08222}{arXiv:1709.08222 [hep-lat]}\BibitemShut
  {NoStop}%
\bibitem [{\citenamefont {Blanton}\ and\ \citenamefont
  {Sharpe}(2020)}]{Blanton:2020jnm}%
  \BibitemOpen
  \bibfield  {author} {\bibinfo {author} {\bibfnamefont {T.~D.}\ \bibnamefont
  {Blanton}}\ and\ \bibinfo {author} {\bibfnamefont {S.~R.}\ \bibnamefont
  {Sharpe}},\ }\href {\doibase 10.1103/PhysRevD.102.054515} {\bibfield
  {journal} {\bibinfo  {journal} {Phys. Rev. D}\ }\textbf {\bibinfo {volume}
  {102}},\ \bibinfo {pages} {054515} (\bibinfo {year} {2020})},\ \Eprint
  {http://arxiv.org/abs/2007.16190}{arXiv:2007.16190 [hep-lat]}\BibitemShut
  {NoStop}%
\bibitem [{\citenamefont {Brice\~no}\ \emph {et~al.}(2017)\citenamefont
  {Brice\~no}, \citenamefont {Hansen},\ and\ \citenamefont
  {Sharpe}}]{Briceno:2017tce}%
  \BibitemOpen
  \bibfield  {author} {\bibinfo {author} {\bibfnamefont {R.~A.}\ \bibnamefont
  {Brice\~no}}, \bibinfo {author} {\bibfnamefont {M.~T.}\ \bibnamefont
  {Hansen}}, \ and\ \bibinfo {author} {\bibfnamefont {S.~R.}\ \bibnamefont
  {Sharpe}},\ }\href {\doibase 10.1103/PhysRevD.95.074510} {\bibfield
  {journal} {\bibinfo  {journal} {Phys. Rev. D}\ }\textbf {\bibinfo {volume}
  {95}},\ \bibinfo {pages} {074510} (\bibinfo {year} {2017})},\ \Eprint
  {http://arxiv.org/abs/1701.07465}{arXiv:1701.07465 [hep-lat]}\BibitemShut
  {NoStop}%
\bibitem [{\citenamefont {Brice\~no}\ \emph {et~al.}(2019)\citenamefont
  {Brice\~no}, \citenamefont {Hansen},\ and\ \citenamefont
  {Sharpe}}]{Briceno:2018aml}%
  \BibitemOpen
  \bibfield  {author} {\bibinfo {author} {\bibfnamefont {R.~A.}\ \bibnamefont
  {Brice\~no}}, \bibinfo {author} {\bibfnamefont {M.~T.}\ \bibnamefont
  {Hansen}}, \ and\ \bibinfo {author} {\bibfnamefont {S.~R.}\ \bibnamefont
  {Sharpe}},\ }\href {\doibase 10.1103/PhysRevD.99.014516} {\bibfield
  {journal} {\bibinfo  {journal} {Phys. Rev. D}\ }\textbf {\bibinfo {volume}
  {99}},\ \bibinfo {pages} {014516} (\bibinfo {year} {2019})},\ \Eprint
  {http://arxiv.org/abs/1810.01429}{arXiv:1810.01429 [hep-lat]}\BibitemShut
  {NoStop}%
\bibitem [{\citenamefont {Hansen}\ \emph {et~al.}(2020)\citenamefont {Hansen},
  \citenamefont {Romero-L\'opez},\ and\ \citenamefont
  {Sharpe}}]{Hansen:2020zhy}%
  \BibitemOpen
  \bibfield  {author} {\bibinfo {author} {\bibfnamefont {M.~T.}\ \bibnamefont
  {Hansen}}, \bibinfo {author} {\bibfnamefont {F.}~\bibnamefont
  {Romero-L\'opez}}, \ and\ \bibinfo {author} {\bibfnamefont {S.~R.}\
  \bibnamefont {Sharpe}},\ }\href {\doibase 10.1007/JHEP07(2020)047} {\bibfield
   {journal} {\bibinfo  {journal} {JHEP}\ }\textbf {\bibinfo {volume} {07}},\
  \bibinfo {pages} {047} (\bibinfo {year} {2020})},\ \bibinfo {note} {[Erratum:
  JHEP 02, 014 (2021)]},\ \Eprint
  {http://arxiv.org/abs/2003.10974}{arXiv:2003.10974 [hep-lat]}\BibitemShut
  {NoStop}%
\bibitem [{\citenamefont {Blanton}\ and\ \citenamefont
  {Sharpe}(2021)}]{Blanton:2020gmf}%
  \BibitemOpen
  \bibfield  {author} {\bibinfo {author} {\bibfnamefont {T.~D.}\ \bibnamefont
  {Blanton}}\ and\ \bibinfo {author} {\bibfnamefont {S.~R.}\ \bibnamefont
  {Sharpe}},\ }\href {\doibase 10.1103/PhysRevD.103.054503} {\bibfield
  {journal} {\bibinfo  {journal} {Phys. Rev. D}\ }\textbf {\bibinfo {volume}
  {103}},\ \bibinfo {pages} {054503} (\bibinfo {year} {2021})},\ \Eprint
  {http://arxiv.org/abs/2011.05520}{arXiv:2011.05520 [hep-lat]}\BibitemShut
  {NoStop}%
\bibitem [{\citenamefont {Blanton}\ \emph {et~al.}(2020)\citenamefont
  {Blanton}, \citenamefont {Romero-L\'opez},\ and\ \citenamefont
  {Sharpe}}]{Blanton:2019vdk}%
  \BibitemOpen
  \bibfield  {author} {\bibinfo {author} {\bibfnamefont {T.~D.}\ \bibnamefont
  {Blanton}}, \bibinfo {author} {\bibfnamefont {F.}~\bibnamefont
  {Romero-L\'opez}}, \ and\ \bibinfo {author} {\bibfnamefont {S.~R.}\
  \bibnamefont {Sharpe}},\ }\href {\doibase 10.1103/PhysRevLett.124.032001}
  {\bibfield  {journal} {\bibinfo  {journal} {Phys. Rev. Lett.}\ }\textbf
  {\bibinfo {volume} {124}},\ \bibinfo {pages} {032001} (\bibinfo {year}
  {2020})},\ \Eprint {http://arxiv.org/abs/1909.02973}{arXiv:1909.02973
  [hep-lat]}\BibitemShut {NoStop}%
\bibitem [{\citenamefont {Hansen}\ \emph {et~al.}(2021)\citenamefont {Hansen},
  \citenamefont {Brice\~no}, \citenamefont {Edwards}, \citenamefont {Thomas},\
  and\ \citenamefont {Wilson}}]{Hansen:2020otl}%
  \BibitemOpen
  \bibfield  {author} {\bibinfo {author} {\bibfnamefont {M.~T.}\ \bibnamefont
  {Hansen}}, \bibinfo {author} {\bibfnamefont {R.~A.}\ \bibnamefont
  {Brice\~no}}, \bibinfo {author} {\bibfnamefont {R.~G.}\ \bibnamefont
  {Edwards}}, \bibinfo {author} {\bibfnamefont {C.~E.}\ \bibnamefont {Thomas}},
  \ and\ \bibinfo {author} {\bibfnamefont {D.~J.}\ \bibnamefont {Wilson}}
  (\bibinfo {collaboration} {Hadron Spectrum}),\ }\href {\doibase
  10.1103/PhysRevLett.126.012001} {\bibfield  {journal} {\bibinfo  {journal}
  {Phys. Rev. Lett.}\ }\textbf {\bibinfo {volume} {126}},\ \bibinfo {pages}
  {012001} (\bibinfo {year} {2021})},\ \Eprint
  {http://arxiv.org/abs/2009.04931}{arXiv:2009.04931 [hep-lat]}\BibitemShut
  {NoStop}%
\bibitem [{\citenamefont {Blanton}\ \emph {et~al.}(2022)\citenamefont
  {Blanton}, \citenamefont {Romero-L\'opez},\ and\ \citenamefont
  {Sharpe}}]{Blanton:2021eyf}%
  \BibitemOpen
  \bibfield  {author} {\bibinfo {author} {\bibfnamefont {T.~D.}\ \bibnamefont
  {Blanton}}, \bibinfo {author} {\bibfnamefont {F.}~\bibnamefont
  {Romero-L\'opez}}, \ and\ \bibinfo {author} {\bibfnamefont {S.~R.}\
  \bibnamefont {Sharpe}},\ }\href {\doibase 10.1007/JHEP02(2022)098} {\bibfield
   {journal} {\bibinfo  {journal} {JHEP}\ }\textbf {\bibinfo {volume} {02}},\
  \bibinfo {pages} {098} (\bibinfo {year} {2022})},\ \Eprint
  {http://arxiv.org/abs/2111.12734}{arXiv:2111.12734 [hep-lat]}\BibitemShut
  {NoStop}%
\bibitem [{\citenamefont {Garofalo}\ \emph {et~al.}(2022)\citenamefont
  {Garofalo}, \citenamefont {Mai}, \citenamefont {Romero-L\'opez},
  \citenamefont {Rusetsky},\ and\ \citenamefont {Urbach}}]{Garofalo:2022pux}%
  \BibitemOpen
  \bibfield  {author} {\bibinfo {author} {\bibfnamefont {M.}~\bibnamefont
  {Garofalo}}, \bibinfo {author} {\bibfnamefont {M.}~\bibnamefont {Mai}},
  \bibinfo {author} {\bibfnamefont {F.}~\bibnamefont {Romero-L\'opez}},
  \bibinfo {author} {\bibfnamefont {A.}~\bibnamefont {Rusetsky}}, \ and\
  \bibinfo {author} {\bibfnamefont {C.}~\bibnamefont {Urbach}},\ }\href@noop {}
  {\  (\bibinfo {year} {2022})},\ \Eprint
  {http://arxiv.org/abs/2211.05605}{arXiv:2211.05605 [hep-lat]}\BibitemShut
  {NoStop}%
\bibitem [{\citenamefont {Bulava}\ \emph {et~al.}(2022)\citenamefont {Bulava},
  \citenamefont {Hansen}, \citenamefont {Hansen}, \citenamefont {Patella},\
  and\ \citenamefont {Tantalo}}]{Bulava:2021fre}%
  \BibitemOpen
  \bibfield  {author} {\bibinfo {author} {\bibfnamefont {J.}~\bibnamefont
  {Bulava}}, \bibinfo {author} {\bibfnamefont {M.~T.}\ \bibnamefont {Hansen}},
  \bibinfo {author} {\bibfnamefont {M.~W.}\ \bibnamefont {Hansen}}, \bibinfo
  {author} {\bibfnamefont {A.}~\bibnamefont {Patella}}, \ and\ \bibinfo
  {author} {\bibfnamefont {N.}~\bibnamefont {Tantalo}},\ }\href {\doibase
  10.1007/JHEP07(2022)034} {\bibfield  {journal} {\bibinfo  {journal} {JHEP}\
  }\textbf {\bibinfo {volume} {07}},\ \bibinfo {pages} {034} (\bibinfo {year}
  {2022})},\ \Eprint {http://arxiv.org/abs/2111.12774}{arXiv:2111.12774
  [hep-lat]}\BibitemShut {NoStop}%
\bibitem [{\citenamefont {Zamolodchikov}\ and\ \citenamefont
  {Zamolodchikov}(1977)}]{Zamolodchikov:1977nu}%
  \BibitemOpen
  \bibfield  {author} {\bibinfo {author} {\bibfnamefont {A.~B.}\ \bibnamefont
  {Zamolodchikov}}\ and\ \bibinfo {author} {\bibfnamefont {A.~B.}\ \bibnamefont
  {Zamolodchikov}},\ }\href {\doibase 10.1016/0550-3213(78)90239-0} {\bibfield
  {journal} {\bibinfo  {journal} {JETP Lett.}\ }\textbf {\bibinfo {volume}
  {26}},\ \bibinfo {pages} {457} (\bibinfo {year} {1977})}\BibitemShut
  {NoStop}%
\bibitem [{\citenamefont {Zamolodchikov}\ and\ \citenamefont
  {Zamolodchikov}(1979)}]{Zamolodchikov:1978xm}%
  \BibitemOpen
  \bibfield  {author} {\bibinfo {author} {\bibfnamefont {A.~B.}\ \bibnamefont
  {Zamolodchikov}}\ and\ \bibinfo {author} {\bibfnamefont {A.~B.}\ \bibnamefont
  {Zamolodchikov}},\ }\href {\doibase 10.1016/0003-4916(79)90391-9} {\bibfield
  {journal} {\bibinfo  {journal} {Annals Phys.}\ }\textbf {\bibinfo {volume}
  {120}},\ \bibinfo {pages} {253} (\bibinfo {year} {1979})}\BibitemShut
  {NoStop}%
\bibitem [{\citenamefont {Guo}(2013)}]{Guo:2013vsa}%
  \BibitemOpen
  \bibfield  {author} {\bibinfo {author} {\bibfnamefont {P.}~\bibnamefont
  {Guo}},\ }\href {\doibase 10.1103/PhysRevD.88.014507} {\bibfield  {journal}
  {\bibinfo  {journal} {Phys. Rev. D}\ }\textbf {\bibinfo {volume} {88}},\
  \bibinfo {pages} {014507} (\bibinfo {year} {2013})},\ \Eprint
  {http://arxiv.org/abs/1304.7812}{arXiv:1304.7812 [hep-lat]}\BibitemShut
  {NoStop}%
\bibitem [{\citenamefont {Brice\~no}\ \emph {et~al.}(2022)\citenamefont
  {Brice\~no}, \citenamefont {Carrillo}, \citenamefont {Guerrero},
  \citenamefont {Hansen},\ and\ \citenamefont {Sturzu}}]{Briceno:2021aiw}%
  \BibitemOpen
  \bibfield  {author} {\bibinfo {author} {\bibfnamefont {R.~A.}\ \bibnamefont
  {Brice\~no}}, \bibinfo {author} {\bibfnamefont {M.~A.}\ \bibnamefont
  {Carrillo}}, \bibinfo {author} {\bibfnamefont {J.~V.}\ \bibnamefont
  {Guerrero}}, \bibinfo {author} {\bibfnamefont {M.~T.}\ \bibnamefont
  {Hansen}}, \ and\ \bibinfo {author} {\bibfnamefont {A.~M.}\ \bibnamefont
  {Sturzu}},\ }\href {\doibase 10.22323/1.396.0315} {\bibfield  {journal}
  {\bibinfo  {journal} {PoS}\ }\textbf {\bibinfo {volume} {LATTICE2021}},\
  \bibinfo {pages} {315} (\bibinfo {year} {2022})},\ \Eprint
  {http://arxiv.org/abs/2112.01968}{arXiv:2112.01968 [hep-lat]}\BibitemShut
  {NoStop}%
\bibitem [{\citenamefont {Wolff}(1989)}]{Wolff:1988uh}%
  \BibitemOpen
  \bibfield  {author} {\bibinfo {author} {\bibfnamefont {U.}~\bibnamefont
  {Wolff}},\ }\href {\doibase 10.1103/PhysRevLett.62.361} {\bibfield  {journal}
  {\bibinfo  {journal} {Phys. Rev. Lett.}\ }\textbf {\bibinfo {volume} {62}},\
  \bibinfo {pages} {361} (\bibinfo {year} {1989})}\BibitemShut {NoStop}%
\bibitem [{\citenamefont {Luscher}\ and\ \citenamefont
  {Wolff}(1990)}]{Luscher:1990ck}%
  \BibitemOpen
  \bibfield  {author} {\bibinfo {author} {\bibfnamefont {M.}~\bibnamefont
  {Luscher}}\ and\ \bibinfo {author} {\bibfnamefont {U.}~\bibnamefont
  {Wolff}},\ }\href {\doibase 10.1016/0550-3213(90)90540-T} {\bibfield
  {journal} {\bibinfo  {journal} {Nucl. Phys. B}\ }\textbf {\bibinfo {volume}
  {339}},\ \bibinfo {pages} {222} (\bibinfo {year} {1990})}\BibitemShut
  {NoStop}%
\bibitem [{\citenamefont {Bulava}(2022)}]{Bulava:2022}%
  \BibitemOpen
  \bibfield  {author} {\bibinfo {author} {\bibfnamefont {J.}~\bibnamefont
  {Bulava}},\ }\href@noop {} {}\bibinfo {howpublished} {{Private
  communication}} (\bibinfo {year} {2022})\BibitemShut {NoStop}%
\bibitem [{\citenamefont {Balog}\ \emph {et~al.}(2009)\citenamefont {Balog},
  \citenamefont {Niedermayer},\ and\ \citenamefont {Weisz}}]{Balog:2009yj}%
  \BibitemOpen
  \bibfield  {author} {\bibinfo {author} {\bibfnamefont {J.}~\bibnamefont
  {Balog}}, \bibinfo {author} {\bibfnamefont {F.}~\bibnamefont {Niedermayer}},
  \ and\ \bibinfo {author} {\bibfnamefont {P.}~\bibnamefont {Weisz}},\ }\href
  {\doibase 10.1016/j.physletb.2009.04.082} {\bibfield  {journal} {\bibinfo
  {journal} {Phys. Lett. B}\ }\textbf {\bibinfo {volume} {676}},\ \bibinfo
  {pages} {188} (\bibinfo {year} {2009})},\ \Eprint
  {http://arxiv.org/abs/0901.4033}{arXiv:0901.4033 [hep-lat]}\BibitemShut
  {NoStop}%
\bibitem [{\citenamefont {Balog}\ \emph {et~al.}(2010)\citenamefont {Balog},
  \citenamefont {Niedermayer},\ and\ \citenamefont {Weisz}}]{Balog:2009np}%
  \BibitemOpen
  \bibfield  {author} {\bibinfo {author} {\bibfnamefont {J.}~\bibnamefont
  {Balog}}, \bibinfo {author} {\bibfnamefont {F.}~\bibnamefont {Niedermayer}},
  \ and\ \bibinfo {author} {\bibfnamefont {P.}~\bibnamefont {Weisz}},\ }\href
  {\doibase 10.1016/j.nuclphysb.2009.09.007} {\bibfield  {journal} {\bibinfo
  {journal} {Nucl. Phys. B}\ }\textbf {\bibinfo {volume} {824}},\ \bibinfo
  {pages} {563} (\bibinfo {year} {2010})},\ \Eprint
  {http://arxiv.org/abs/0905.1730}{arXiv:0905.1730 [hep-lat]}\BibitemShut
  {NoStop}%
\end{thebibliography}%
